\documentclass[sigconf]{acmart}

\renewcommand\footnotetextcopyrightpermission[1]{} % removes footnote with conference information in first column
\usepackage{booktabs} % For formal tables
\usepackage{subcaption}

\acmConference[iPres2018]{International Conference on Digital Preservation}{September 2018}{Boston and Cambridge, Massachusetts USA}
\acmYear{2018}
\copyrightyear{2018}

\begin{document}

\title{The Many Shapes of Archive-It}

%\titlenote{Produces the permission block, and
%  copyright information}
\subtitle{Characteristics of Archive-It Collections}
%\subtitlenote{The full version of the author's guide is available as
%  \texttt{acmart.pdf} document}

%\author{(withheld for reviewing)}
%\orcid{(withheld for reviewing)}
%\affiliation{%
%	\institution{(withheld for reviewing)}
%	\city{(withheld for reviewing)}
%	\state{(withheld for reviewing)}
%	\postcode{(withheld for reviewing)}
%}
%\email{(withheld for reviewing)}
%
%\author{(withheld for reviewing)}
%\orcid{(withheld for reviewing)}
%\affiliation{%
%	\institution{(withheld for reviewing)}
%	\city{(withheld for reviewing)}
%	\state{(withheld for reviewing)}
%	\postcode{(withheld for reviewing)}
%}
%\email{(withheld for reviewing)}
%
%\author{(withheld for reviewing)}
%\orcid{(withheld for reviewing)}
%\affiliation{%
%	\institution{(withheld for reviewing)}
%	\city{(withheld for reviewing)}
%	\state{(withheld for reviewing)}
%	\postcode{(withheld for reviewing)}
%}
%\email{(withheld for reviewing)}
%
%\author{(withheld for reviewing)}
%\orcid{(withheld for reviewing)}
%\affiliation{%
%	\institution{(withheld for reviewing)}
%	\city{(withheld for reviewing)}
%	\state{(withheld for reviewing)}
%	\postcode{(withheld for reviewing)}
%}
%\email{(withheld for reviewing)}

\author{Shawn M. Jones}
%\authornote{Dr.~Trovato insisted his name be first.}
\orcid{1234-5678-9012}
\affiliation{%
  \institution{Old Dominion University}
%  \streetaddress{P.O. Box 1212}
  \city{Norfolk}
  \state{Virginia}
  \postcode{23529}
}
\email{sjone@cs.odu.edu}

\author{Alexander Nwala}
\affiliation{%
  \institution{Old Dominion University}
%  \streetaddress{P.O. Box 1212}
  \city{Norfolk}
  \state{Virginia}
  \postcode{23529}
}
\email{anwala@cs.odu.edu}

\author{Michele C. Weigle}
%\authornote{This author is the
%  one who did all the really hard work.}
\affiliation{%
  \institution{Old Dominion University}
%  \streetaddress{P.O. Box 1212}
  \city{Norfolk}
  \state{Virginia}
  \postcode{23529}
}
\email{mweigle@cs.odu.edu}

\author{Michael L. Nelson}
%\authornote{The secretary disavows any knowledge of this author's actions.}
\affiliation{%
  \institution{Old Dominion University}
%  \streetaddress{P.O. Box 1212}
  \city{Norfolk}
  \state{Virginia}
  \postcode{23529}
}
\email{mln@cs.odu.edu}

%\affiliation{%
% \institution{Rajiv Gandhi University}
%\affiliation{%
%  \institution{Tsinghua University}
%  \streetaddress{30 Shuangqing Rd}
%  \country{China}
%\affiliation{%
%  \institution{Palmer Research Laboratories}
%  \streetaddress{8600 Datapoint Drive}
%\affiliation{\institution{The Th{\o}rv{\"a}ld Group}}
%\email{jsmith@affiliation.org}
%
%\author{Julius P.~Kumquat}
%\affiliation{\institution{The Kumquat Consortium}}
%\email{jpkumquat@consortium.net}

\renewcommand{\shortauthors}{S. Jones et al.}

\begin{abstract}
%Web archives meet the needs of journalists, social scientists, historians, and government organizations. The use cases for these groups often require that they guide the archiving process themselves, creating their own web archive collections. Archive-It is a subscription service started by the Internet Archive in 2005 for the purpose of allowing organizations to create their own web archive collections. Understanding these collections could be done via their user-supplied metadata or via text analysis. Instead, our work proposes using structural metadata as an additional way to understand these collections. We adapt the concept of the collection growth curve for understanding Archive-It collection curation and crawling behavior. We also introduce several seed features to describe the diversity and types of seeds present in an Archive-It collection. Finally, we use the descriptions of each collection to identify four semantic categories of Archive-It collections. Using the identified structural features, we are able to predict the semantic category of a collection using a Random Forest classifier with a weighted average $F_1$ score of 0.720, thus bridging the structural to the descriptive.
Web archives, a key area of digital preservation, meet the needs of journalists, social scientists, historians, and government organizations. The use cases for these groups often require that they guide the archiving process themselves, selecting their own original resources, or seeds, and creating their own web archive collections. We focus on the collections within Archive-It, a subscription service started by the Internet Archive in 2005 for the purpose of allowing organizations to create their own collections of archived web pages, or mementos. Understanding these collections could be done via their user-supplied metadata or via text analysis, but the metadata is applied inconsistently between collections and some Archive-It collections consist of hundreds of thousands of seeds, making it costly in terms of time to download each memento. Our work proposes using structural metadata as an additional way to understand these collections. We explore structural features currently existing in these collections that can unveil curation and crawling behaviors. We adapt the concept of the collection growth curve for understanding Archive-It collection curation and crawling behavior. Using the growth curves, we can see if most of the mementos in the collection are skewed earlier or later. We also introduce several seed features to describe the diversity and types of seeds present in an Archive-It collection. With these seed features, we come to an understanding of the diversity of resources that make up a collection and the depth of those resources within their seed websites, indicating whether the curator chose to preserve the top-level page or something more specific within a site. Finally, we use the descriptions of each collection to identify four semantic categories of Archive-It collections. Using the identified structural features, we reviewed the results of runs with 20 classifiers and are able to predict the semantic category of a collection using a Random Forest classifier with a weighted average $F_1$ score of 0.720, thus bridging the structural to the descriptive. Our method is useful because it saves the researcher time and bandwidth. They do not need to download every resource in the collection in order to identify its semantic category. Identifying collections by their semantic category allows further downstream processing to be tailored to these categories.
\end{abstract}




\maketitle

\section{Introduction}

Web archiving has become an important area of digital preservation as news, research, and other content publishing has moved to the web. Government organizations seek to archive their web presence for posterity \cite{uknatarchives}. Historians \cite{milligan2016}, social scientists \cite{MEET:MEET14504801096, arms2006}, and journalists \cite{ohlheiser2017, hafner2017} use web archives to understand human behavior.  Because libraries have a focus on curating content specific to their communities, web archiving was once identified as a ``growth area for library collections'' \cite{chudnov2011}. In order to facilitate the creation of curated collections, the Internet Archive created Archive-It in 2005 \cite{mcclure2006}. Archive-It is a subscription service with the goal of allowing organizations and users to create their own web archive collections. Archive-It collections, in particular, are interesting because a single organization is responsible for each collection, meaning that the curation strategy for a collection is guided by humans rather than automated crawling operations. As web archives, Archive-It collections reflect changes to individual resources over time, providing a chronicle of unfolding world events, or the history of an organization. These changes over time make Archive-It collections different from more traditional document collections which typically contain only one version of a given resource.

How do we understand an Archive-It collection? We could look at its descriptive metadata, effectively asking others what they have said about it. We could evaluate the URI of each item in the collection to locate and then download its contents, thus \textbf{dereferencing} each URI to produce content for analysis.  Such analysis can use techniques such as text mining, effectively looking at the collection's parts. We will show that other sources of information exist to provide the shapes necessary to understand the collection. What behaviors exist in web archive collections that we cannot acquire merely from metadata or text analysis? What structural features exist that can unveil curation and crawling behaviors? In this work, we examine structural features and determine what shapes exist within Archive-It collections. With each shape, we gain a better understanding of the collection.

The point of this work is to demonstrate what we can be learned through the structural features of Archive-It collections. Using only structural metadata is advantageous because it saves one from having to dereference all of the URIs in a collection in order to understand it. Some collections have hundreds of thousands of mementos, such as the \emph{2014 Primaries} collection with 2380 seeds and 219,084 mementos of those seeds. Our prior work summarized Archive-It collections by selecting representative items from them. These representative items were then visualized using social media storytelling \cite{AlNoamany:2017:GSA:3091478.3091508}. That work focused mainly on collections centered on events. In this work, we identify other types of collections that may have been overlooked by our previous summarization efforts.

For a given collection, we seek to answer questions about its temporal nature. Does most of the collection exist earlier or later in its life? When did the archivist select and archive a collection's contents? Did they create a collection at start that was intended to archive new versions of the same web pages repeatedly in perpetuity? Did they nurture the selected web pages throughout the collection's life and add content continuously? Was there renewed interest at some point later in the collection's life? To answer those questions we adapt the concept of growth curves, first introduced by AlSum \cite{AlSum2014}, to Archive-It collections.

We also seek to answer questions about the web pages selected for the collection. Was the collection built from web sites belonging to one organization or many? Were most of the web pages in the collection top-level pages or specific articles deeper in a web site? To answer these questions we introduce concepts like domain diversity and seed path depth diversity.

%To what semantic categories can we map these structural features?  

Furthermore, how well can we bridge the structural to the descriptive? We will discuss how Archive-It collections fit into four different semantic categories. As noted above, our prior work only focused on collections based on events, the smallest category. How well can we use our structural features to classify a collection into one of these semantic categories? Using the RandomForest classifier from Weka \cite{datraminingbook, frank2016} we show that we can predict these semantic categories with a weighted average $F_1$ of 0.720. 

We believe that the creation of these semantic categories, as well as how well our structural features predict them, makes this work a unique contribution, because one can use these structural features to infer meaning without having to dereference all of the web pages in an Archive-It collection. This is useful because one can use this identification to support further, more specific processing tailored to that semantic category.

\section{Background}

Archive-It collections consist of archived web pages, or \textbf{mementos}. An archivist creates these mementos from a list of URIs known as \textbf{seeds}, also known as ``Original Resources''. Thus seed selection is the genesis of a collection. The archivist then instructs a crawler \cite{archiveit-crawling} to create mementos from these seeds at certain intervals. The crawler produces a new version of each seed with each crawl. Depending on the chosen configuration, an archivist can also instruct the crawler to visit any additional web pages linked to from the seeds. This process produces even more mementos. To reduce confusion in this work, we will use the term \textbf{seed memento} to specifically refer to mementos created from seeds. Seed mementos are of particular interest because they tie back to decisions made explicitly by the archivist and thus represent unique policy and behavior for each collection.

\begin{figure}[t]
\centering
\includegraphics[scale=0.18]{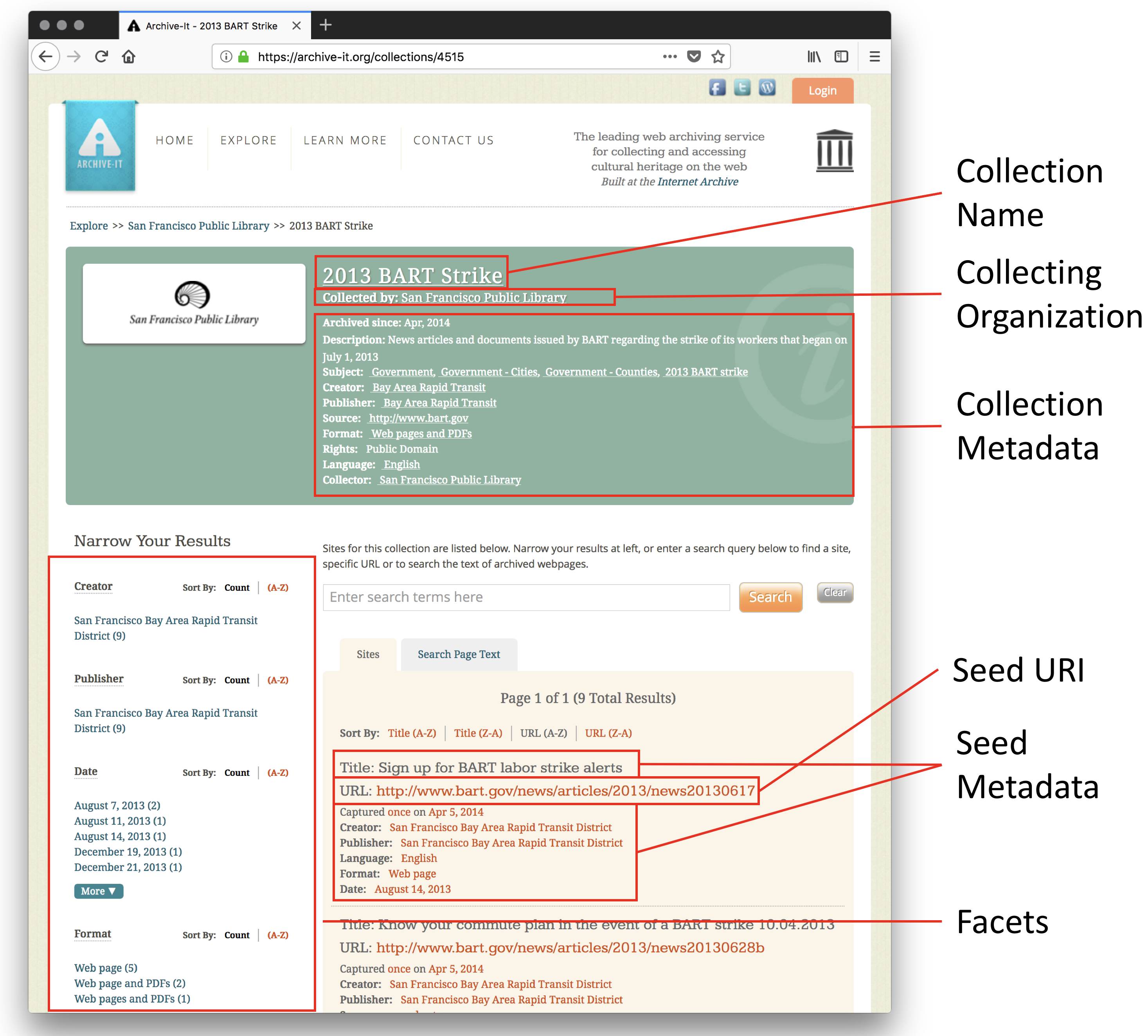}
\caption{Example Collection from \url{https://archive-it.org/collections/4515}}
\label{fig:archiveit_screenshot}
\end{figure}

Archive-It's user interface for each collection can provide the viewer a wealth of information, as shown in Figure \ref{fig:archiveit_screenshot}, including a description, collection metadata, seed metadata, and facets for discovery and searching. The archivist is only required to supply the collection name. The collecting organization in the \emph{collected by} field is drawn from the archivist's account. Collections can be marked public, in which seed URIs and seed metadata is available, or private, where only collection metadata is available. Each collection includes an \emph{archived since} field, indicating when the collection was created.

With the exception of the collection name, the collecting organization, and the \emph{archived since} field, all additional metadata is optional and provided by the archivist. For each collection, the archivist can choose from a controlled vocabulary of fields based on Dublin Core \cite{archiveit-metadata} or they can use their own freeform vocabulary. These same fields can be applied individually for each seed. Archivists may also select one or more topics for the collection. Like the metadata fields, there is a controlled vocabulary, but archivists may add additional topics as well. Unfortunately, this metadata is inconsistently applied to collections, likely due to differences in content standards and rules of interpretation among archivists.

Being compliant with the Memento Protocol \cite{rfc7089}, each Archive-It collection also provides a \textbf{TimeMap} for each seed. TimeMaps are lists of mementos for a specific seed. Using TimeMaps, one can acquire the URIs of all mementos for a given seed as well as each memento's \textbf{memento-datetime}, the datetime that the memento was recorded.

\section{Related Work}

Digital collections, and more specifically, those at Archive-It, have been explored by others in the past.

Fenlon explores the different types of digital collections that exist \cite{PRA2:PRA214505401010}. She mentions that a digital collection's contents, metadata, and even user interface are constructed based on the needs of the audience that they serve. Even though she did not focus on web archives, her work is related because it indicates that there is additional appetite among scholars to understand the features of digital collections outside of our summarization efforts.

Milligan \cite{Milligan:2016:CSC:2910896.2910913} reviews three web archive collections to determine the effectiveness of different techniques for choosing seeds. The three collections differed in how seeds were chosen: (1) through seed URIs extracted from tweets within a given Twitter hashtag, (2) from general crawling via the Internet Archive, or (3) manually by curators at the University of Toronto Libraries. He discovered that a combination of hashtags and careful curation proves best. It is the behavior of those who create this third type of collection that we study in this paper. Likewise, Nwala evaluates how to use search engine result pages (SERPs) to discover news stories appropriate for building or augmenting web archive collections \cite{nwala2018}. Our work differs because we analyze the seeds after selection.

Work has been done to understand the behaviors of the users who create collections with live web curation platforms. Using the Uses and Gratifications model \cite{doi:10.1111/j.00117315.2004.02524.x}, Mull \cite{MULL2014192} discovered the following motivations for using the image curation platform Pinterest: ``fashion'', ``creative projects'', ``entertainment'', ``virtual exploration'', and ``organization''.  Wang \cite{doi:10.1177/2056305116662173} applied the MAIN model \cite{sundar2008main} to explain the different gratifications of Pinterest users in an attempt to understand why users engaged with the platform. The results of Wang's study indicate that Pinterest users create collections for the purpose of engaging with the topics that they find to be fun and exciting, in pursuit of escapism. The users analyzed in these studies are different from the institutions that create collections in Archive-It. Those institutions have different motivations for creating a collection. Some have legal requirements to archive government agencies. Others collect resources on behalf of scholars at the institution. Our work involves understanding what behavior can be derived from the features in web archive collections rather than conducting user studies to understand their motivations for creating collections.

%Wang \cite{} and Mull \cite{} by apply Uses and Gratifications (U\&G) theory \cite{} to understand the motivations of Pinterest users. 

%Wang explores the different gratifications of Pinterest users in an attempt to understand why they use the different features of the platform. 

Ogden brings to light the behavior and work of web archivists \cite{Ogden:2017:OWA:3091478.3091506}. She applies an ethnographic approach to understand those who participate in the work of the Internet Archive, noting that they are currently focused on methods for discovering seeds.  Crook used Archive-It as part of an effort to produce a web archive of online Australian publications \cite{doi:10.1108/02640470910998542}. He highlights the challenges of using the Archive-It service, especially for those used to having complete control over the archiving and playback processes. Slania \cite{doi:10.1086/669993} and Deutch \cite{doi:10.1086/685975} detail the challenges of using Archive-It to archive art web sites. Where their work focuses on the impressions of web archivists, ours focuses on studying the output of their work. We review their behavior as observed from the structural features of Archive-It collections themselves.

Niu evaluated the capabilities of ten different web archives, including Archive-It, highlighing features such as keyword searching and date facets \cite{niu2012}. Rather than criticizing or evaluating the capabilities of Archive-It, our work is intended to highlight structural features that may be used to better understand its collections.

Encoded Archival Description (EAD) \cite{eadsite} is an ``XML standard for encoding archival finding aids'' maintained by the Library of Congress. Archive-It does not currently use EAD, instead favoring a metadata scheme based on Dublin Core \cite{dublincore}. Our work attempts to identify structural features that exist within web archive collections rather than relying upon existing metadata.

% remove or redact?
%Alam profiled web archives by using summarization of CDX files available from some web archives \cite{Alam2016}. In that work, the authors focused on extracting information from web archives with the intent of improving routing within aggregation services. Archive-It was one of the corpora for which they produced results. Our work instead presents seed features and archiving behaviors not yet explored for understanding web archive collections.

AlNoamany evaluated different methods of detecting off-topic pages within web archives, focusing on Archive-It \cite{AlNoamany2016ijdl}. Sa\u{g}lam sought to use the content of specific Archive-It collections to analyze the timeliness of medical data through the use of information retrieval techniques \cite{SAGLAM2014104}. We are looking at structural features rather than the content of the web archive collections themselves.

Abramson focuses on machine learning techniques that can classify URIs based on their contents without dereferencing (downloading) them \cite{AAAIW125252}. Though it could be used to augment our summarization work, we currently focus on other aspects of URIs and their diversity within a collection.

AlSum analyzes different web archives to determine which seeds they cover over which periods of time \cite{AlSum2014}. To acquire seeds, the authors randomly sampled URIs from DMOZ as well as search engines provided by the web archives themselves. AlSum discusses the age of the mementos within each archive, which top-level domains are covered, which languages exist within the mementos of the archive, and the growth curves of each corpus over time. We use these growth curves in our own work. We also review all seeds and seed mementos available within Archive-It, focusing on the growth curves at a collection level rather than at the level of an entire web archive.

Like this work, AlNoamany also reviews some characteristics of archived collections within Archive-It \cite{AlNoamany2016}. That analysis, performed in 2015, focused on the number of seeds, the mean number of mementos per seed, the time span of each collection, the domains within these collections, which top level domains were represented, and the decay rate of resources within all Archive-It collections. Our work is different because we look at collections as units unto themselves and have developed different metrics to measure them.

\section{Data Acquisition}

Our data breakdown for this work is shown in Table \ref{tab:data_breakdown}. To acquire seed URIs, we used BeautifulSoup \cite{Nair:2014:GSB:2616225, beautifulsoupdocs} to scrape the web pages of 9,351 Archive-It collections between October and December of 2017. Because private collections do not expose seeds, we were only able to acquire the public Archive-It collections. For each public collection there also exists a comma-separated report of seed URIs. In 6\% of collections, the number of seeds acquired between web scraping and this report did not match. To account for these cases, we used the union of these two sources to get the list of seeds for each collection.

We did not use the Archive-It CDX/C API because it requires knowing seed URIs beforehand \cite{archiveit-cdxc} and we did not use the optional OAI-PMH interface available for many collections \cite{archiveit-oaipmh} because it would not provide all of the information we needed. Also, not all collections have enabled this feature.

Once we had the seeds, we extracted domain names using tldextract \cite{Kurkowski2017}. We acquired the seed memento URIs and their memento-datetimes via Memento TimeMaps. Because they are effectively empty, we removed collections from our dataset that had no mementos. Some of these TimeMaps were empty due to downloading or processing errors, and thus their collections were removed. We also removed all collections containing \textbf{singletons}, consisting of a single seed with a single memento, because they do not provide enough information to create growth curves. Likewise, we removed collections where all mementos were crawled within the same second. For semantic analysis, we reviewed each collection's name and associated metadata and removed collections that were marked by the archivist with the terms \emph{test} or \emph{trial}. This left us with 3,382 collections consisting of 700,835 seeds and 6,943,677 seed mementos to review.

\begin{table}[t]
\centering
\caption{Reduction of Data for this Paper}
\label{tab:data_breakdown}
\begin{tabular}{p{4cm} r r}
\textbf{Data Category Description}                            & \textbf{Count} & \textbf{Remaining} \\\hline
Total Collections                               & 9,351           &                    \\\hline
Removed Private Collections                     & 4,823           & 4,528               \\\hline
Removed Collections Archived Since Jan, 2017    & 440            & 4,088               \\\hline
Removed Collections \\ With No Mementos            & 248            & 3,840               \\\hline
Removed Collections \\ With Errors	& 	48	& 3,792 \\ \hline
Removed Singletons				& 	357		&  3,435		\\ \hline
Removed Single Second \\Collections 	& 21	&	3,414 \\ \hline
Removed Test Collections			& 	32		&  3,382		\\\hline
Collections Remaining For Analysis &                & 3,382      
\end{tabular}
\end{table}

\section{Structural Features}

\subsection{Collection Growth Curves}

\begin{figure}
\centering
\includegraphics[scale=0.25]{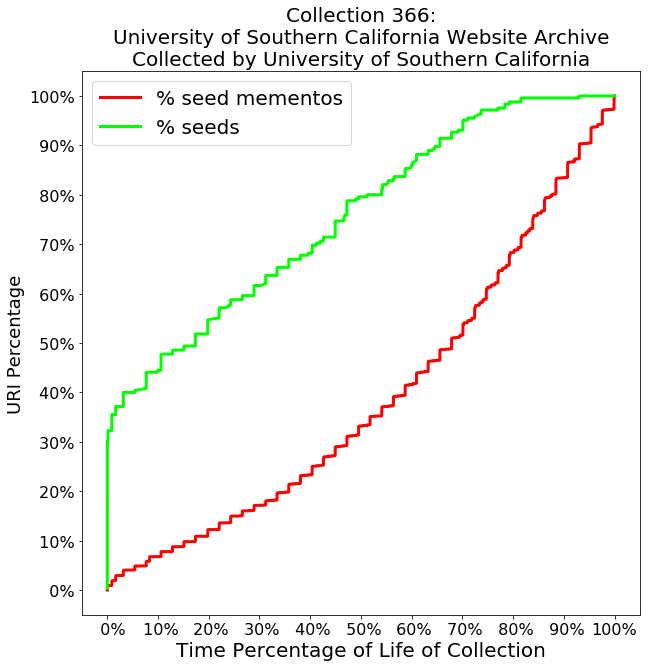}
\caption{The Growth Curve of Archive-It Collection 366}
\label{fig:example_growth_curve}
\end{figure}

A collection growth curve provides insight into the seed curation and crawling behavior of an Archive-It collection. Figure \ref{fig:example_growth_curve} shows an example growth curve for Archive-It Collection 366\footnote{Archive-It collections can be accessed by appending the collection number to the URI https://archive-it.org/collections/, so collection 366 would be \url{https://archive-it.org/collections/366}}. The x-axis represents the \textbf{life of the collection}, or the time between a collection's first memento and its last. To normalize among collections with different durations, we show the x-axis as a percentage of the collection's lifespan. The y-axis represents the percentage of URIs in the collection at a given time. URIs come in two categories: seeds or seed mementos, represented by the green and red lines, respectively.

\begin{figure}
\centering
\includegraphics[scale=0.2]{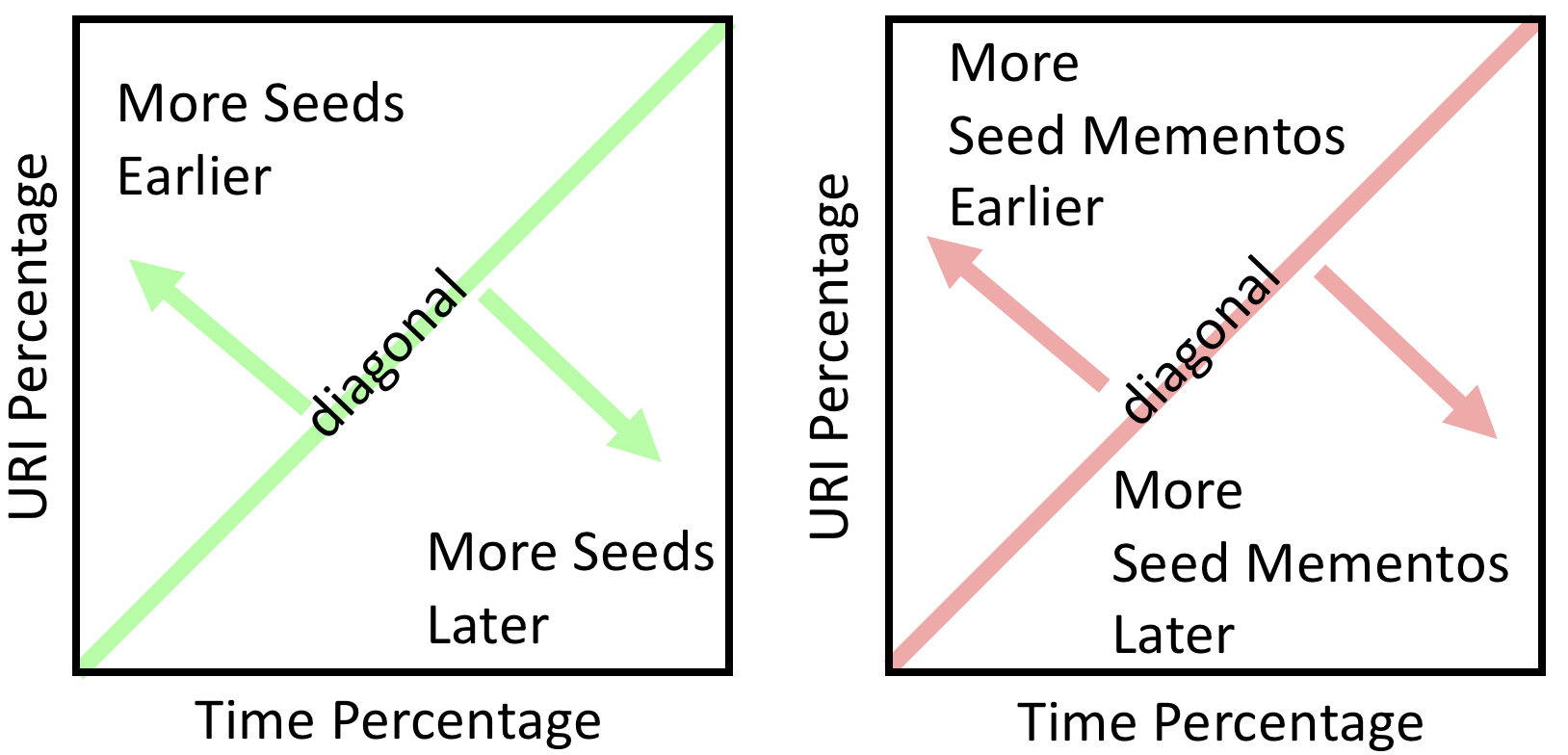}
\caption{The Anatomy of a Collection Growth Curve}
\label{fig:growth_curve_anatomy}
\end{figure}

Growth curves for Archive-It collections consist of multiple parts. Figure \ref{fig:growth_curve_anatomy} demonstrates how to interpret the information within a growth curve. An imaginary diagonal line shows a linear relationship between the growth of URIs over time. It divides each graph into two parts. If the seed line (green) is in the upper left corner, then most of the seeds were added earlier to the collection, and if the seed line occupies the lower right corner, then most were added later. The seed line reflects an aspect of curatorial engagement with the collection, indicating when the archivist first crawled a given seed. The closer the seed line is to the diagonal, the more often the archivist added a new seed. The memento-datetime of the first memento for each seed is used to generate the seed line.

\begin{figure*}
	\centering
	\begin{subfigure}{0.3\textwidth}
	\centering
	\captionsetup{justification=centering}
	\includegraphics[scale=0.22]{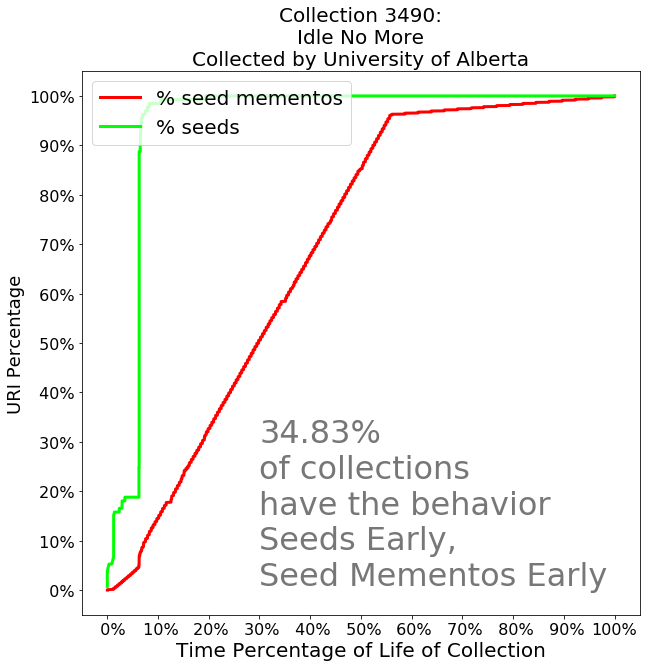}
	\caption{\\Archive-It Collection 3490: \\Seeds early,\\seed mementos early}
	\label{fig:seme_example}
	\end{subfigure}%
	\begin{subfigure}{0.3\textwidth}
	\centering
	\captionsetup{justification=centering}
	\includegraphics[scale=0.22]{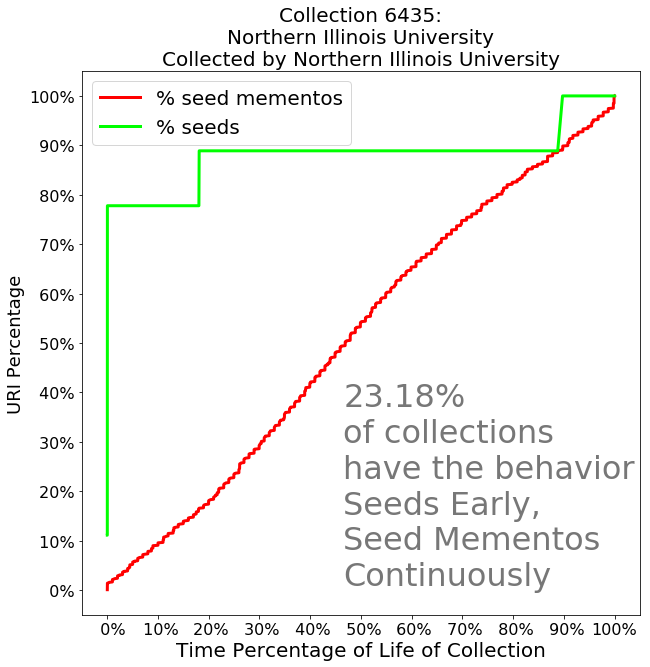}
	\caption{\\Archive-It Collection 6435: \\Seeds early, \\seed mementos continuously}
	\label{fig:semd_example}
	\end{subfigure}%
	\begin{subfigure}{0.3\textwidth}
	\centering
	\captionsetup{justification=centering}
	\includegraphics[scale=0.22]{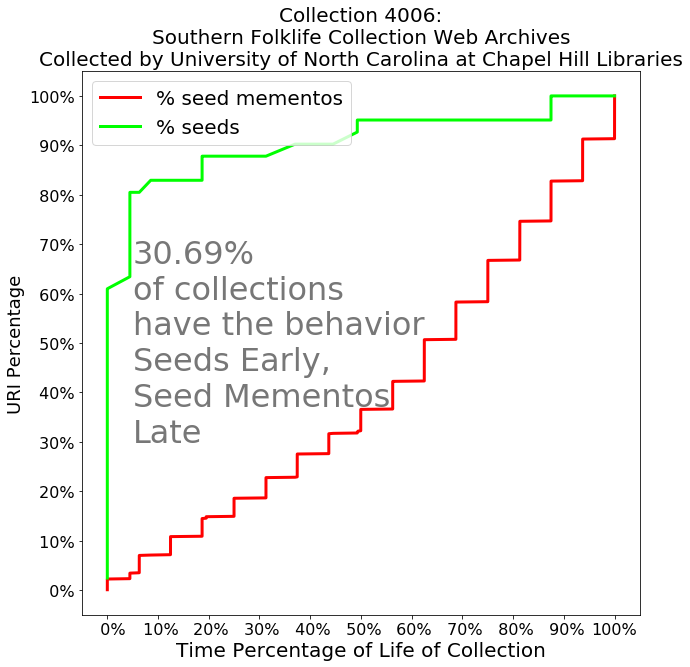}
	\caption{\\Archive-It Collection 4006: \\Seeds early, \\seed mementos late}
	\label{fig:seml_example}
	\end{subfigure}
	
	\begin{subfigure}{0.3\textwidth}
	\centering
	\captionsetup{justification=centering}
	\includegraphics[scale=0.22]{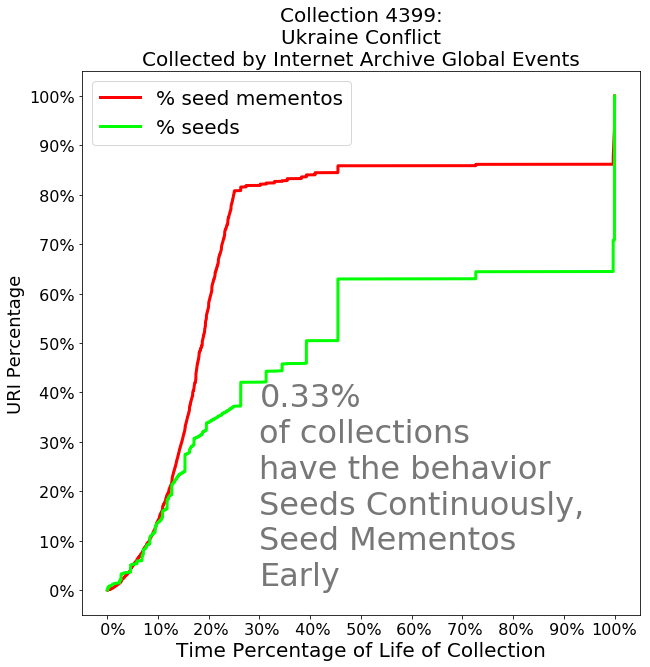}
	\caption{\\Archive-It Collection 4399:\\Seeds continuously, \\seed mementos early}
	\label{fig:sdme_example}
	\end{subfigure}%
	\begin{subfigure}{0.3\textwidth}
	\centering
	\captionsetup{justification=centering}
	\includegraphics[scale=0.22]{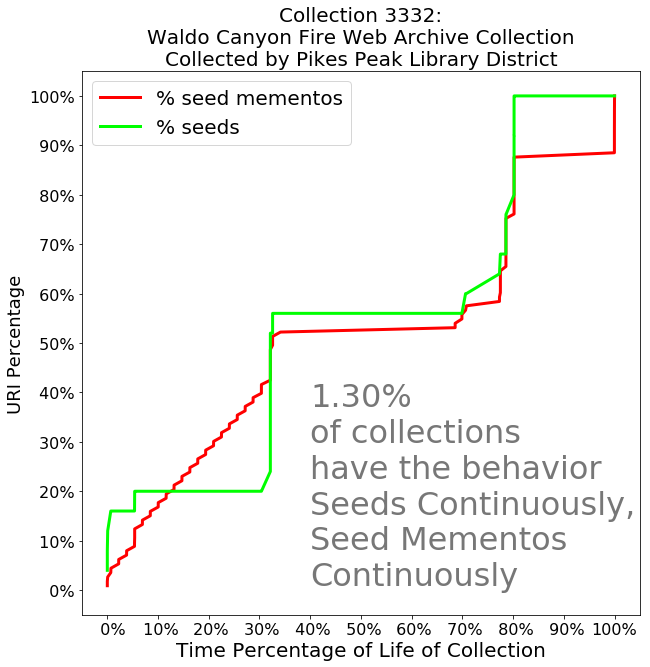}
	\caption{\\Archive-It Collection 3332:\\Seeds continuously, \\seed mementos continuously}
	\label{fig:sdmd_example}
	\end{subfigure}%
	\begin{subfigure}{0.3\textwidth}
	\centering
	\captionsetup{justification=centering}
	\includegraphics[scale=0.22]{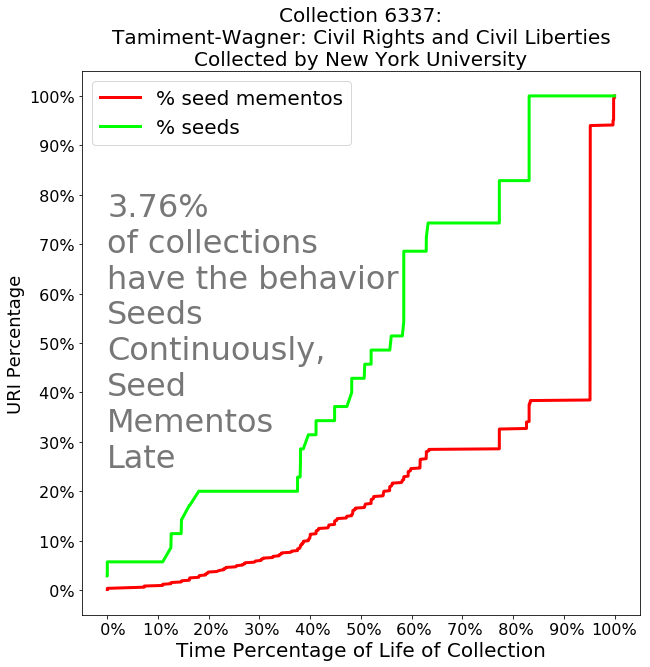}
	\caption{\\Archive-It Collection 6337:\\Seeds continuously, \\seed mementos later}
	\label{fig:sdml_example}
	\end{subfigure}
	
	\begin{subfigure}{0.3\textwidth}
	\centering
	\captionsetup{justification=centering}
	\includegraphics[scale=0.22]{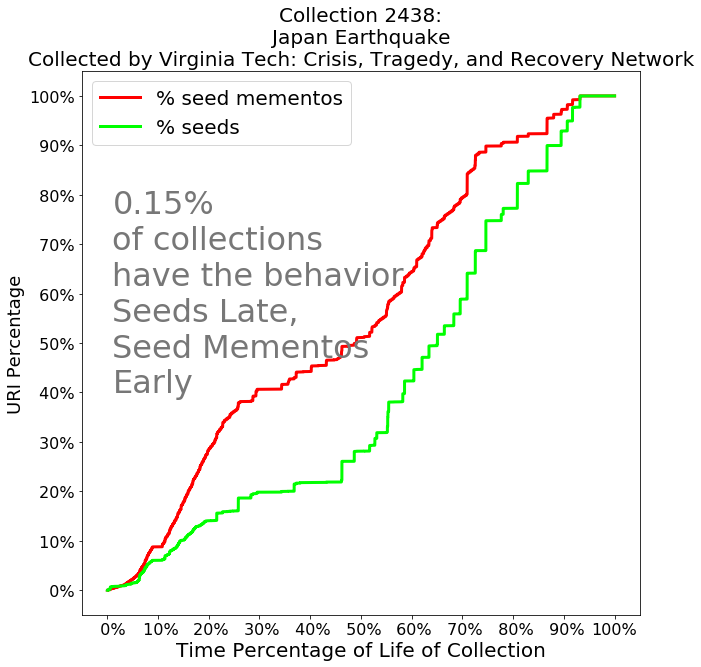}
	\caption{\\Archive-It Collection 2438:\\Seeds Late, \\seed mementos early }
	\label{fig:slme_example}
	\end{subfigure}%
	\begin{subfigure}{0.3\textwidth}
	\centering
	\captionsetup{justification=centering}
	\includegraphics[scale=0.22]{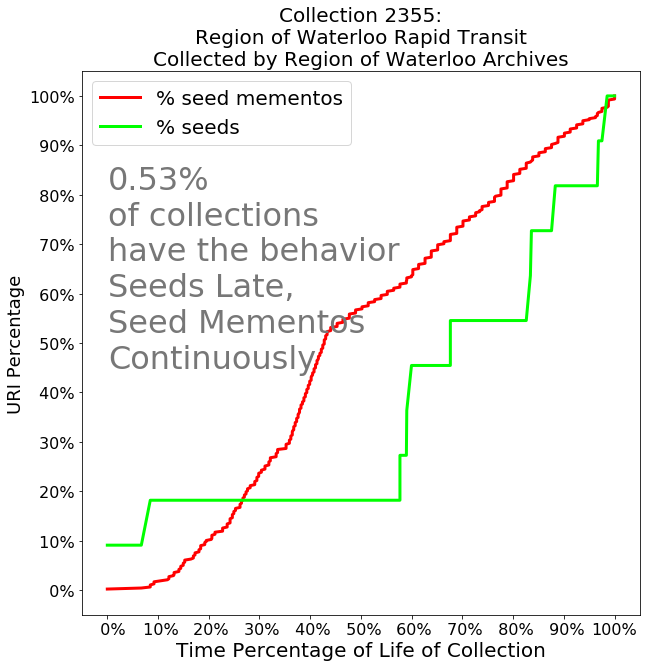}
	\caption{\\Archive-It Collection 2355:\\Seeds late, \\seed mementos continuously}
	\label{fig:slmd_example}
	\end{subfigure}%
	\begin{subfigure}{0.3\textwidth}
	\centering
	\captionsetup{justification=centering}
	\includegraphics[scale=0.22]{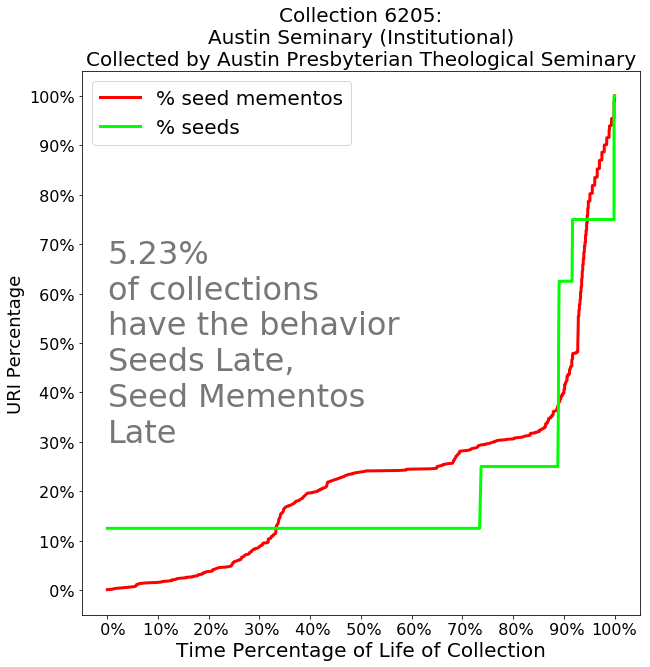}
	\caption{\\Archive-It Collection 6205:\\Seeds late, \\seed mementos late}
	\label{fig:slml_example}
	\end{subfigure}
\caption{Examples from nine different growth curve behavior categories, grey inset text conveys the percentage of the 3,382 collections in this study}
\label{fig:growth_curve_classification}
\end{figure*}

The meaning of the seed memento line is similar. Where the seed line indicates intent, the seed memento line indicates the growth of actual collection. If the seed memento line (red) mostly occupies the upper left corner, then most of the mementos were crawled earlier in the collection's life, meaning that most of the collection's holdings were created at that time, and its temporality is skewed earlier. If the seed memento line occupies the lower right corner, then the collection's temporality is skewed later. If the seed memento line runs along the diagonal, then the collection's temporality is spread more evenly across the collection. The memento-datetimes of all mementos are used to generate the seed memento line.

\subsubsection{Interesting Growth Curve Behaviors}

Using the area under the curve (AUC), we were able to identify different collection behaviors. If the AUC exceeds 0.5 --- the area of the diagonal --- then the growth was earlier. If the AUC is less than 0.5 then the growth was later. If the AUC is within 0.05 of the diagonal, then we considered it growing \emph{continuously}. Using these three primitives, we identified the behaviors shown in Figure \ref{fig:growth_curve_classification}.

\textbf{Seeds early, seed mementos early} --- Seen in Figure \ref{fig:seme_example} with collection \emph{Idle No More} (ID 3490), the growth curves with this behavior indicate that the archivist made most curatorial decisions near the start of the life of the collection. The seed memento line is skewed early, indicating that most of the seed mementos in the collection come from that time period.

\textbf{Seeds early, seed mementos continuously} --- Figure \ref{fig:semd_example} shows collection \emph{Northern Illinois University} (ID 6435), where the archivist added more than 70\% of the seeds near the beginning of the collection's life. The archivist selected seeds early, but then chose crawling strategies that added seed mementos steadily throughout its existence. 

\textbf{Seeds early, seed mementos late} ---  In all of these cases, the archivist chose seeds early, but the crawling strategy produced seed mementos at a later time. In collection \emph{Southern Folklife Collection Web Archives} (ID 4006), shown in Figure \ref{fig:seml_example}, we see a case where the archivist added 60\% of the seeds earlier in the collection's life. The crawling strategy ensured that seed mementos were crawled later. In this case 50\% of all mementos were crawled by 65\% of its life. 

Seeds early is the most frequent seed behavior, taking place in 88.7\% of all collections studied.

%\begin{figure*}[t]
%\centering
%\begin{subfigure}{0.4\textwidth}

%\begin{figure}[t]
%\centering
%\captionsetup{justification=centering}
%\includegraphics[scale=0.22]{all_up_front_example.png}
%\caption{\\Archive-It Collection 1169: \\all seeds up front behavior}
%\label{fig:example_growth_all_seeds_up_front}
%\end{figure}

%\end{subfigure}%
%\begin{subfigure}{0.4\textwidth}
%\centering
%\captionsetup{justification=centering}
%\includegraphics[scale=0.22]{overlapping_example.png}
%\caption{\\Archive-It Collection 1909: \\curve overlapping behavior}
%\label{fig:example_curve_overlapping}
%\end{subfigure}
%\label{fig:other_curve_behaviors}
%\caption{Other curve behaviors}
%\end{figure*}

\textbf{Seeds continuously, seed mementos early} --- 
Figure \ref{fig:sdme_example} shows collection \emph{Ukraine Conflict} (ID 4399), where the seed growth curve wraps around the diagonal, indicating that the archivist added seeds more regularly, but most crawling happened earlier in the collection's life. This means that there are more seed mementos from the earlier seeds. 

\textbf{Seeds continuoulsy, seed mementos continuously} --- Collection \emph{Waldo Canyon Fire Web Archive Collection} (ID 3332) is shown in Figure \ref{fig:sdmd_example}. Collections with this behavior indicate continuous involvement both on the part of seed selection as well as crawling. Both lines wrap the diagonal as the collection grows steadily. 

\textbf{Seeds continuously, seed mementos late} --- Shown in Figure \ref{fig:sdml_example} with collection \emph{Tamiment-Wagner: Civil Rights and Civil Liberties} (ID 6337), this behavior indicates that the archivist was continuously engaged in adding seeds to the collection, but most of the mementos were created later in the collection's life. 

\textbf{Seeds late, seed mementos early} --- This behavior is demonstrated by collection \emph{Japan Earthquake} (ID 2438) in Figure \ref{fig:slme_example}. In this case, the seed memento growth line exists farther left on the graph than the seed growth line. The early seeds added to the collection have more memento growth than the seeds that follow, because the archivist added more seeds later in the collection's life.

\textbf{Seeds late, seed mementos continuously} --- Figure \ref{fig:slmd_example} shows collection \emph{Region of Waterloo Rapid Transit} (ID 2355). In this case, the seed memento growth is steady, but something changed around 60\% of its life span. Approximately 60\% of the seed mementos belong to the first 20\% of seeds.

\textbf{Seeds late, seed mementos late} --- This behavior is exemplified by collection \emph{Austin Seminary (Institutional)} (ID 6205), shown in Figure \ref{fig:slml_example}. In this case, more seeds were added when the collection was already 70\% old. Another dramatic shift happened when more seeds were added at the 90\% mark and again later. This could indicate dramatically renewed interest in this collection. 

% Seed mementos late is the most frequent seed memento behavior, taking place in 39.68\% of all collections studied.

\textbf{Memento growth overtakes seed growth} --- Sometimes the seed growth stops for a while and the seed memento growth overtakes it, visualized as the seed memento line being higher than the seed line. Figures \ref{fig:sdme_example}, \ref{fig:slme_example}, \ref{fig:slmd_example}, and \ref{fig:slml_example} all demonstrate this behavior.

\textbf{All seeds up front} --- All of the seeds in these collections were chosen at the beginning of the collection's life, but the seed memento growth varies. In these cases, the archivist not only chose the seeds early in the collection's life, but they never added seeds later. It is an extreme case of the seed line from Figure \ref{fig:seme_example}.

\textbf{Curve overlapping} --- Sometimes the ratio of seeds to seed mementos is 1-to-1 or very close to 1-to-1. The lines in these cases show up as overlapping. Their growth behavior indicates that very few crawls are happening per seed. In these cases, it is likely that the archivist just wanted to gather a single copy of a given seed rather than recording the changes to a seed over time. Because we removed singletons from the dataset, all of these collections have more than one seed.

\subsubsection{Growth Curve Features}

We have identified five growth curve features which provide insight into the behavior of a collection.

The \textbf{number of seeds} submitted to the collection varies, as does the \textbf{number of seed mementos}. These can be counted by using the seed acquisition activities and TimeMaps mentioned above.

\textbf{Difference between seed curve AUC and diagonal} --- The AUC of the seed curve indicates whether the seeds were added earlier or later to the collection. Subtracting this value from the AUC of the diagonal gives additional information useful for understanding the nature of the seed curve. Negative values indicate that seeds were added later. Positive values indicate that seeds were added earlier. Values very close to 0 indicate that seeds were added continuously.

\textbf{Difference between seed memento curve AUC and diagonal} -- The area under the seed memento curve is useful as well. Subtracting this value from the AUC of the diagonal provides similar information to the seed AUC feature mentioned above.

\textbf{Difference between seed curve AUC and seed memento curve AUC} --- Subtracting the AUC of both curves indicates how close they are to each other. A value of 0 indicates that the curves are overlapping, likely meaning that there is one memento per seed. A positive value means that the seeds are added earlier than the seed mementos. A negative value means that the seed memento growth has overtaken the seed growth.

\textbf{Collection Lifespan} --- The collection lifespan is the difference in memento-datetime between the last memento and the first.

% \textbf{Number of Seed Mementos} --- There are a number of reasons for which a collection contains a large number of seed mementos. Some collections have a high number of seed mementos due to a high collection lifespan. Other collections 

\subsection{Seed Features}

In addition to the growth curves, structural features of the seeds themselves provide insight into the behavior of the archivist with respect to a collection.

\textbf{Seed URI domain diversity} --- Seed URI domain diversity \cite{nwala_blog} quantifies the spread of the collection across different sources. A collection where all seeds are from the same domain would have a domain diversity of 0 and one where all seeds are from different domains would have a domain diversity of 1. Equation \ref{eq:diversity} computes the diversity $D$ as the number of unique domains $U$ divided by the number of seeds $C$. In Equation \ref{eq:diversity_normalized} $D'$ normalizes this diversity value $D$ between 0 and 1. A collection with 1 seed, by definition, has a diversity of 0.

\begin{equation} \label{eq:diversity}
D = \frac{U}{C}
\end{equation}

\begin{equation} \label{eq:diversity_normalized}
D' = \frac{CD - 1}{C - 1} = \frac{U - 1}{C - 1}
\end{equation}

%\textbf{Seed URI domain diversity without social media} --- We created an alternative to domain diversity by removing all social media domains from the domain diversity calculations. Some collections are archiving the web presence of an entity, and the additional social media domains provide noise to a collection that would otherwise have a domain diversity of 0.

%We separated the domains from the rest of the seed URI using the tldextract library\footnote{\url{https://pypi.python.org/pypi/tldextract}}. Figure \ref{fig:domain_diversity} displays the domain diversity for all Public Archive it collections. Only 10.1\% of collections have a domain diversity of $0$, indicating that their seed URIs come from the same domain. Another 30.1\% have a domain diversity between $0$ and $0.1$ meaning that most of the collection consists of seed URIs from the same domain. This accounts for the apparent horizontal line across the bottom of the plot. Because single seeds all have a domain diversity of $0$, we include Figure \ref{fig:domain_diversity_no_single_seeds} which shows the domain diversity with the 1,241 single seed collections removed. On the other end of spectrum 18.1\% of collections have a domain diversity of $1$, indicating that all seed URIs in the collections come from different domains.

The \textbf{path depth} for each seed URI consists of the number of items separated by slashes after the domain name. If the path consists of a query string, a 1 is added to the path depth, similar to  \cite{mccown2005}. If the last item in the path consists of a known default page (e.g., index.html), then we subtract 1 from the path depth. Default pages are determined by a list of well known default pages\footnote{\url{https://support.tigertech.net/index-file-names}}. Path depths indicate crawling intention by the archivist with respect to the collection. Recall that seeds are the starting point for a crawl, thus an archivist who selects a path depth of 0 seeks to start crawling from the top of a web site, whereas one who starts with a higher path depth may be starting with a page containing more specific content.

\textbf{Seed URI path depth diversity} --- We acquire an idea of the spread of path depth across the collection by applying the above domain diversity equation to the seed path depth of every seed in the collection. This may indicate if the seed URIs consist solely of top-level pages or a mixture of top-level pages and more specific content.

%Comparing path depth diversity to domain diversity shows that 3.2\% of all collections have a path depth diversity of $1$ and a domain diversity of $1$. In these cases, every seed URI has a different domain name and all path depths differ. These collections contain between $2$ and $5$ seeds.

%At the other extreme 30.4\% of collections have seeds with a path depth diversity of $0$ and a domain diversity of $0$, meaning that these collections have seeds from the same domain and all at the same path depth. All of these collections consist of a single seed.

\textbf{Most frequent seed URI path depth} --- If a collection's most frequent seed URI path depth is 0, then it mostly consists of seeds of web site top-level pages. If the most frequent path depth is higher, then it mostly consists of seeds deeper in a web site.

\textbf{\% Query string usage} --- Some collections consist mostly of URIs with query strings, whereas others consist of just paths. A collection with 1 seed has either 0\% or 100\% query string usage.

%For each collection, we calculated the percentage of seeds in which a given path depth occurred. The path depth with the highest percentage is the \textbf{most frequent path depth}. Figure \ref{fig:most_frequent_path_depths} shows the most frequent path depths compared with the number of collections with that value. Here we see that the most frequent path depth is indeed 0, meaning that most public Archive-It collections contain seed URIs of the top level URI for a web site.

\section{Bridging the Structural to the Descriptive}
%
%\begin{table}[t]
%\centering
%\caption{Distribution of collections for each semantic category}
%\label{tab:distribution-semantic-categories}
%\begin{tabular}{l r r}
%\textbf{Semantic} & \textbf{\# of} & \textbf{\% of All} \\
%\textbf{Category} & \textbf{Collections} & \textbf{Collections} \\ \hline
%Self-Archiving & 1,832 &  53.8\% \\ \hline
%Subject-based Archiving & 943 & 27.7\%  \\  \hline
%Time Bounded - Expected  & 485 &  14.3\% \\ \hline
%Time Bounded - Spontaneous & 143 & 4.2\% \\ \hline
%\hline
%Total & 3403 & 100\% \\ \hline
%\end{tabular}
%\end{table}

\begin{table}[t]
\centering
\small
\caption{Distribution of collections for each semantic category}
\label{tab:distribution-semantic-categories}
\begin{tabular}{l r r}
\textbf{Semantic} & \textbf{\# of} & \textbf{\% of All} \\
\textbf{Category} & \textbf{Collections} & \textbf{Collections} \\ \hline
Self-Archiving & 1,828 &  54.1\% \\ \hline
Subject-based Archiving & 935 & 27.6\%  \\  \hline
Time Bounded - Expected  & 476 &  14.1\% \\ \hline
Time Bounded - Spontaneous & 143 & 4.2\% \\ \hline
\hline
Total & 3,382 & 100\% \\ \hline
\end{tabular}
\end{table}

Each structural feature tells part of the story of a collection. We also wanted to know how well these features mapped to the descriptive metadata for each collection. Our goal is to be able to predict some aspect of the descriptive information from the structural features introduced in the last section.

Every Archive-It collection has one or more assigned topics. Some of these topics come from a controlled vocabulary, but the archivist has the option of providing one or more freeform topics of their own as well. We evaluated several classifiers to predict these controlled vocabulary topics. The best weighted average $F_1$ score we achieved was 0.225 with the Logistic Model Tree classifier \cite{Landwehr2005}. The poor results were likely due to the one-to-many relationship of these archivist-assigned topics. 

Considering that the user-supplied topics were not suitable for prediction, we reviewed the Archive-It collections by hand and placed them into four \textbf{semantic categories}. The distribution of these semantic categories is shown in Table \ref{tab:distribution-semantic-categories}.

\textbf{Self-Archiving} ---  These collections consist of one or more domains either (1) belonging to the archiving organization, or (2) being archived as part of some archiving initiative of which the collecting agency is part. Collections fitting into this category include the \emph{University of Utah Web Archive} (ID 2278) archived by the University of Utah, or the \emph{City of Eagan Websites} (ID 2289) archived by the City of Eagan, Minnesota. In each case the organization is archiving its own web presence. Less apparent are collections like \emph{Governor of Tennessee, Phil Bredesen } (ID 391) archived by the Tennessee State Library and Archives. In these latter cases, the archiving organization, the name of the collection, and/or the ownership of the seeds do not match, but the Tennessee State Library and Archives specifically exists to archive the State of Tennessee government. Tennessee State Libraries has collections for many, if not all, Tennessee state agencies. From this behavior, we can infer that they are tasked with archiving the web presence of all Tennessee state government. Other organizations with collections that fit into this category are the Federal Depository Library Program Web Archive and the Region of Waterloo Archives.

\textbf{Subject-based Archiving} --- Some collections consist of a number of seeds bound by a single topic. The topic may be clear, as with \emph{Environmental Justice} (ID 7635), archived by Tufts University. The topic may also be vague, like \emph{ISU Special Collections Department Manuscript Collections Web Sites} (ID 1501) archived by Iowa State University (ISU). In the former, the subject is in the collection name. In the latter the subject is organizations that have shared physical items with the ISU Special Collections Department. This is not Self-Archiving because these organizations are not part of ISU, nor is it apparent that ISU is specifically tasked via some broader archiving initiative to archive them. ISU is merely complementing their physical library collection by archiving additional information about the organizations who have contributed to it.

% Subject-based Archiving also breaks down further into two more special-case categories, bound by time.

\textbf{Time Bounded - Expected} --- These collections focus on an expected, planned event, such as \emph{2008 Olympics} (ID 871) archived by the University of North Carolina. The collections may also be based on a specific time period, such as \emph{Virginia's Political Landscape, 2007} (ID 663) archived by the Library of Virginia. Collections from institutions participating in the K-12 Archiving Initiative \cite{k12archiving} also fit into this category, as they are planned to exist for a single semester or school year.

\textbf{Time Bounded - Spontaneous} --- These collections start after a spontaneous event. Collections fitting into this category include \emph{Tucson Shootings} (ID 2305) archived by the Virginia Tech: Crisis, Tragedy, and Recovery Network and \emph{2011 Japan Earthquake} (ID 2450) archived by the University of Michigan, School of Information. They may also start after the beginning of a movement, such as  \emph{Black Lives Matter Movement} (ID  6396) archived by the San Jose State University, School of Information. The key is that these collections are started due to this spontaneous event or movement and are usually terminated at some point.

%\subsection{Prediction Results}

We determined that Random Forest \cite{Breiman2001} is the best classifier for predicting the semantic category. We arrived at this conclusion by testing 20 classifiers with Weka v. 3.8.2 \cite{frank2016}. We set the semantic category as the target class and evaluated several machine learning algorithms with 10-fold cross validation. Table \ref{tab:top10-classifiers} shows the weighted average results of classification runs using these structural features for the top 10 classifiers we evaluated. We have four classes, so these weighted average results do not provide a complete picture. Table \ref{tab:random-forest-by-class} shows the results for Random Forest by semantic category. Self-archiving scores the best, with an weighted average $F_1$ score of 0.847. This is likely due to the fact that 54.1\% of Archive-It collections fall into this category. Other semantic categories do not fare so well. Time Bounded - Spontaneous is the worst, with $F_1$ of 0.456. This is likely because it only makes up 4.2\% of all collections, giving the classifier little with which to train. More surprising is that Time Bounded - Expected does so well at 0.621, even though it only makes up 14.1\% of all collections. Finally, Subject-based Archiving is slightly worse with an $F_1$ of 0.562 in spite of making up 27.6\% of all collections. Random Forest also had the best $F_1$ score of all classifiers in all semantic categories. Thus, Random Forest trained on our data set performs best at identifying to what semantic category a collection belongs.

Better results for almost all semantic categories can be achieved by removing the number of mementos feature. We converted each of the four semantic categories to a numeric value of $1$ - $4$  and then calculated Kendall's $\tau$ on the feature-category combination, with the results shown in Table \ref{tab:feature-semantic-correlation}.  Domain diversity and collection lifespan have the highest correlation to the categories, with scores of 0.3863 and -0.3320, respectively.  The number of mementos and the most frequent URI depth have lowest correlation. We removed the lowest-scoring attributes one at a time. Removing the number of mementos feature and retraining with Random Forest, shown in Table \ref{tab:random-forest-without-number-of-mementos}, improved the $F_1$ scores of Subject-based Archiving, Time Bounded - Expected, and Time Bounded - Spontaneous to 0.568, 0.641, and 0.462 respectively. The score of Self-Archiving went down from 0.847 to 0.755. Removing more features and retraining did not improve the $F_1$ scores further.

\begin{table}[t]
\centering
\small
\caption{Weighted average results of 10 best classifiers for predicting semantic class evaluated using 10-fold cross validation runs while training on the complete data set}
\label{tab:top10-classifiers}
\begin{tabular}{l r r r r}
                      & \multicolumn{4}{c}{\textbf{Weighted Average}}                        \\
\textbf{Classifier}   & \textbf{TPR} & \textbf{FPR} & \textbf{$F_1$} & \textbf{ROC Area} \\ \hline
Random Forest         & 0.728        & 0.182        & 0.720               & 0.871             \\ \hline
ForestPA              & 0.713        & 0.201        & 0.701              & 0.854             \\ \hline
Decision Table        & 0.702        & 0.214        & 0.685              & 0.831             \\ \hline
LMT                   & 0.702        & 0.205        & 0.685              & 0.833             \\ \hline
CDT                   & 0.700          & 0.212        & 0.686              & 0.819             \\ \hline
JRip                  & 0.698        & 0.235        & 0.679              & 0.769             \\ \hline
Simple Cart           & 0.694        & 0.199        & 0.683              & 0.811             \\ \hline
FT                    & 0.693        & 0.201        & 0.681              & 0.789             \\ \hline
BFTree                & 0.689        & 0.214        & 0.676              & 0.766             \\ \hline
Multilayer Perceptron & 0.686        & 0.197        & 0.675              & 0.818
\end{tabular}
\end{table}

\begin{table}[t]
\small
\centering
\caption{Results by class of Random Forest classifier for predicting semantic category evaluated using 10-fold cross validation runs}
\label{tab:random-forest-by-class}
\begin{tabular}{l r r r r}
                      & \multicolumn{4}{c}{\textbf{Weighted Average}}   \\
\textbf{Semantic Category} & \textbf{TPR} & \textbf{FPR} & \textbf{$F_1$} & \textbf{ROC Area} \\ \hline
Self-Archiving             & 0.891        & 0.250         & 0.847              & 0.899             \\ \hline
Subject-based Archiving    & 0.538        & 0.144        & 0.562              & 0.794             \\ \hline
Time Bounded \\ - Expected    & 0.588        & 0.050         & 0.621              & 0.911             \\ \hline
Time Bounded \\ - Spontaneous & 0.364        & 0.010         & 0.456              & 0.879             \\ \hline
Weighted Average           & 0.728        & 0.182        & 0.720               & 0.871            
\end{tabular}
\end{table}

\begin{table}[t]
\centering
\small
\caption{Kendall $\tau$ Correlation of Features to Semantic Categories}
\label{tab:feature-semantic-correlation}
\begin{tabular}{p{6cm} r}
\textbf{Feature}                                             & \textbf{Kendall $\tau$} \\ \hline
Seed URI domain diversity                                        & 0.3863               \\ \hline
Collection lifespan                                          & -0.3320              \\ \hline
Number of seeds                                              & 0.2878               \\ \hline
Difference between seed curve AUC and seed memento curve AUC & -0.2416              \\ \hline
\% query string usage                                        & 0.2265               \\ \hline
Difference between seed memento Curve AUC and diagonal       & 0.1569               \\ \hline
Difference between seed curve AUC and diagonal               & -0.1387              \\ \hline
Seed URI path depth diversity                                    & 0.1317               \\ \hline
Most frequent seed URI path depth                                      & -0.0687              \\ \hline
Number of mementos                                           &  0.0561
\end{tabular}
\end{table}

%\begin{table}[t]
%\centering
%\small
%\caption{Results by class for Random Forest classifier for predicting semantic category with Time Bounded categories combined}
%\label{tab:random-forest-by-class-time-bounded-combined}
%\begin{tabular}{l r r r r}
%                      & \multicolumn{4}{c}{\textbf{Weighted Average}}   \\
%\textbf{Semantic Category}       & \textbf{TPR}   & \textbf{FPR}   & \textbf{F-Measure} & \textbf{ROC Area} \\ \hline
%Self-Archiving          & 0.889 & 0.242 & 0.849     & 0.897    \\ \hline
%Subject-based Archiving & 0.509 & 0.130  & 0.551     & 0.795    \\ \hline
%Time Bounded            & 0.669 & 0.062 & 0.687     & 0.920     \\ \hline
%Weighed Avg.            & 0.744 & 0.178 & 0.737     & 0.873   
%\end{tabular}
%\end{table}

\begin{table}[t]
\centering
\small
\caption{Results by class for Random Forest classifier with the number of mementos feature removed}
\label{tab:random-forest-without-number-of-mementos}
\begin{tabular}{l r r r r}
                      & \multicolumn{4}{c}{\textbf{Weighted Average}}   \\
\textbf{Semantic Category}       & \textbf{TPR}   & \textbf{FPR}   & \textbf{$F_1$} & \textbf{ROC Area} \\ \hline
Self-Archiving          			& 0.881 	& 0.251 	& 0.841	& 0.891    \\ \hline
Subject-based Archiving 		& 0.549 	& 0.146  	& 0.568	& 0.782    \\ \hline
Time Bounded - Expected		& 0.609 	& 0.048 	& 0.641	& 0.906     \\ \hline
Time Bounded - Spontaneous 	& 0.364 	& 0.009	& 0.462	& 0.877	\\ \hline
Weighed Average            		& 0.729 	& 0.183 	& 0.722	& 0.862   
\end{tabular}
\end{table}

\section{Future Work}

As noted in the introduction, we intend to adapt these structural features and our classifier results to better support our collection summarization work \cite{AlNoamany:2017:GSA:3091478.3091508}. For example, if growth curves indicate that a collection's mementos are skewed earlier, we can select different mementos for our storytelling summarization. The seed analysis features provide information on the diversity of a collection, allowing us to change our algorithms to better choose which seeds are included. Using the classifier, we can tailor summarization algorithms to specific semantic categories of Archive-It collections. We can also augment these features with semantic information as well, such as by analyzing the seed URIs with Abramson's method \cite{AAAIW125252}.

Although we were able to extract most of our needed data using Memento and screen scraping, a structured metadata API would have been helpful. We intend to work further with Archive-It to develop this capability via the WASAPI project \cite{wasapi-presentation} or similar work.

\section{Conclusions}

Archive-it collections can be understood, but not only via their metadata or contents. We have shown that there exist structural features that provide additional information on the shapes necessary to understand a collection. In addition to the number of seeds and number of seed mementos, more complex features exist.

We have adapted collection growth curves to Archive-It collections, revealing their behaviors. Collection growth curves consist of two lines, a seed line and a seed memento line. The seed line indicates when a seed was first added to a collection. The seed memento line conveys the crawling behavior for all seeds in a collection over time. Using these two curves, we can see if the archivist controlling a collection added seeds early, late, or continuously, indicating the level of curatorial involvement with the collection. Likewise, we can see if the seed mementos were crawled early, late, or continuously, indicating the crawling strategy of a collection. We discovered that most collections have their seeds skewed early. Through these curves we gain an understanding of the skew of the temporality of a collection.

We have also identified seed features that help with understanding the curation strategy of a collection. Via domain diversity, we can tell if a collection consists of seeds from one domain or many, thus understanding that the collection comes from many different sources. Using the most frequent URI path depth, we determine if most of the collection consists of top-level pages or specific deep URIs. With seed path depth diversity, we understand the spread of path depths within a collection, indicating if most of its seeds have the same path depth. Understanding how much of a collection uses a query string in its URIs also provides information on the nature of its seeds. Through these features, we gain an understanding of the nature of what was chosen for archiving.

We bridged the structural and the descriptive by classifying collections into four semantic categories: Self-Archiving, Subject-based Archiving, Time Bounded - Expected, and Time Bounded - Spontaneous. We discovered that Self-Archiving is the most prevalent semantic category, making up 54.1\% of the collections surveyed. We also provided the results of training runs with classifiers, and determined that the Random Forest classifier performs best at identifying the semantic category, with a weighted average $F_1$ score of 0.720. We discussed how one could improve the scores of the classifier for most semantic categories by removing the number of mementos feature.

In this work, we have shown that semantic metadata and the collection's holdings themselves are not the whole picture and that there are many shapes to Archive-It.

\begin{acks}
This work supported in part by the Institute of Museum and Library Services (LG-71-15-0077-15).
\end{acks}

%\clearpage
\bibliographystyle{ACM-Reference-Format}
\bibliography{bibliography}
%\bibliography{double-blind-bibliography}

\end{document}